# Managing shop floor systems Using RFID in the Middle East

Dr. Adel Ghannam

and

Dr.Salwa Nassar,Dr. Hesham El Deeb

## Abstract

The controllability of a factory is highly dependent on the capability of a given enterprise planning system to interact with the shop floor, and the capability of any authorized user to review the operation plans , as well as the status of any sales order on the shop floor. We have called the concept " E-manufacturing" ,which is a concept to improve the controllability of a factory by connecting the Enterprise Resource Planning (ERP) system to the shop floor control (SFC) system through the Internet . Business systems and plant systems must be coupled to reduce decision cycle times and increase plant productivity , as well as eliminating human intervention to improve accuracy and data availability speed.

This paper is the output of a project funded by the Egyptian Ministry of Scientific Research and Technology, through the Electronic Research Institute . The **overall objective** of the ministerial program is bridging the gap between R&D and manufacturing to fulfil the immediate technological needs of the Egyptian industry and economy , and building a national technology know-how expertise . This is achieved, in this project , by addressing the development of an E-Manufacturing system using available RFID state-of-the-art technology . **The specific objective** is developing an end product for Egyptian industry that manages real-time interaction with back end ERP systems to correct delays on the shop floor, improving real-time operation , reducing cost, and allow real-time visibility , through the intelligent usage of the RFID.The end product is a fully working system tested in the factory of the manufacturing partner MOBICA ( http://mobica.net/Index.html )

This report addresses the logical and technical structure ,as well as the technology constraints that should be considered in such systems, and the implementation stages of the project

## Problem and challenge background

For the Enterprise-resources-planning system (ERP), every work center is a data point, which is the place where a specific type of data needs to be read or a control data needs to be injected. In the Middle East, many firms are performing this task manually using paper documents (dispatch list, work orders, maintenance orders…etc). E-manufacturing is the name we have given to the concept of improving the controllability of a factory by online electronically connecting the Enterprise Resource Planning (ERP) system to the plant operations usually called Shop-floor-control.

Back end Business systems (ERP) and plant systems must be coupled to reduce decision time and improve the production plan accuracy. In other words, the basic focus of any coupling approach – is Work-in-process (WIP) monitor and control. Fig.1 shows how the ME industry currently provides this coupling in a traditional manual way, compared to the E.Manufacturing concept. The result is that the daily ERP planning output lacks the actual status that should be considered besides the sales orders and marketing forecasts, as well management lacks real-time visibility into WIP on the shop floor. Real time visibility is necessary in the target manufacturing Make-To-Order and Engineer-to-Order

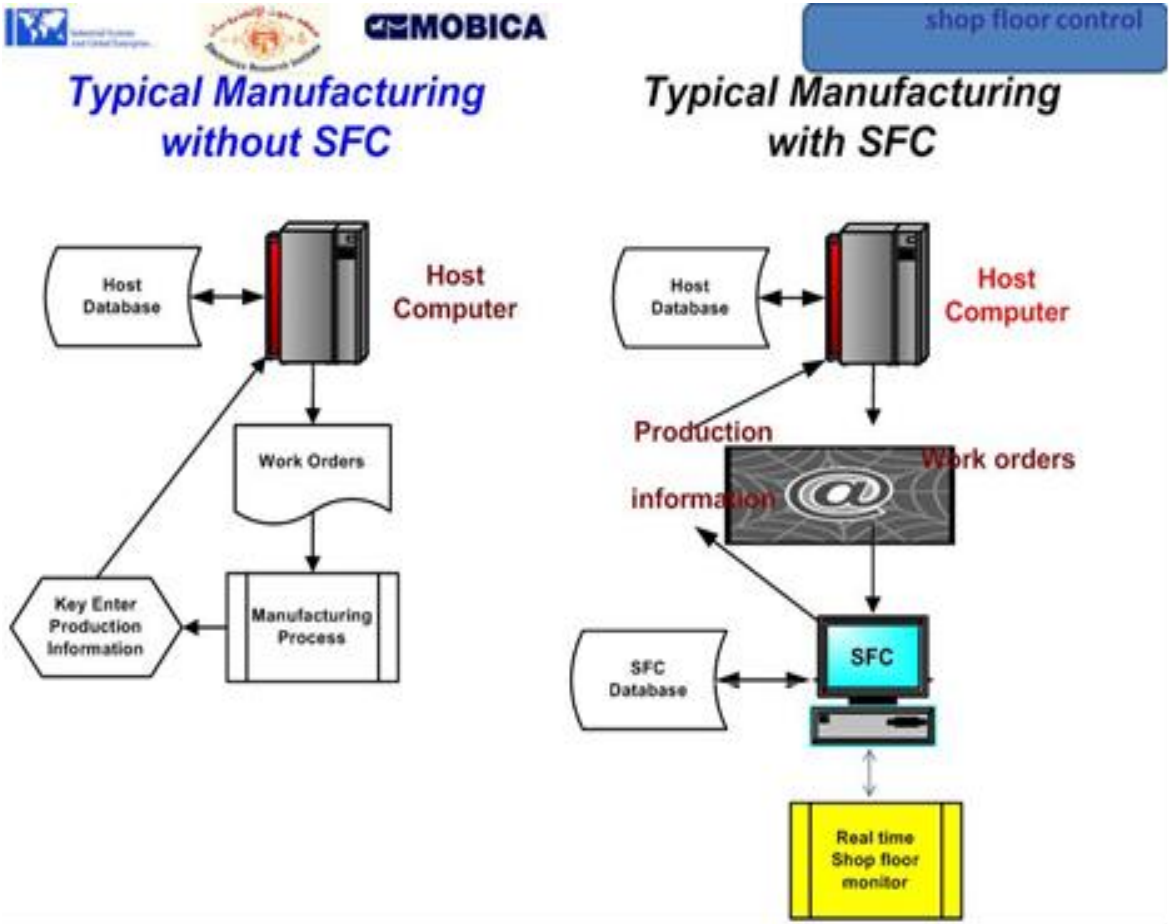

Fig.1-The Shop floor control (SFC) coupling with ERP

Moreover, on-line connection of the Shop floor control to the internet allows any user to inquire the status of any order on the shop floor. The user could be an internal employee or a manager from the enterprise located in any place of the world. This real-time visibility is essential for companies that manufacture-to-sales –order and Engineer-to-order. . Factories adopting this manufacturing strategy needs to compete on the edge of high-mix, low-volume, custom-engineered manufacturing environments and low cost . This manufacturing strategy is highly recommended nationally in Egypt , because the advanced developed industrial countries already excel in the low-mix ,high volume make-to-stock . Egypt needs to excel in a different manufacturing strategy .

## Information flow

A basic element in the system is the data point .A data point is the location selected to fix the RFID reader .The work-In-process( WIP ) is monitored through a build ticket tag that moves along the routing path , starting from the input gate of material, as defined in the Dispatch list imported from the ERP using web services. With the WIP tag we track the progress in real-time on the floor through readers at each work center .At any instant , we have a full visibility of the WIP data across all processes and stations through readers at each work center ( the WIP data points )

Each build tag contains the following fields:

a. The customer sales order, or the internal order in case of Make-to-stock
b. The output product ID
c. The route ID ,which identifies the sequence of operations at work centers. In job-shop manufacturing we may have parallel routes for the product. The limited memory of low cost tags ( 512 bits ) imposes that data required while moving the tag from one station to the next ,such as detailed routing , can be obtained interactively from the dispatch list on the SFC server. Routing is important to make the system viable for Job-shop manufacturing .

Each work center on the shop floor is monitored by a set of data reader + antenna, see Fig.2 ( some time more than one antenna is required ).The set is hanged at an altitude of 5 mt and can read any RFID tag in the input buffer area of the monitored work center. The input buffer area is a horizontal floor's circle having a radius of 3.6 mt. The data point is also equipped with a touch screen monitor displaying the WIP tags waiting in the input buffer area

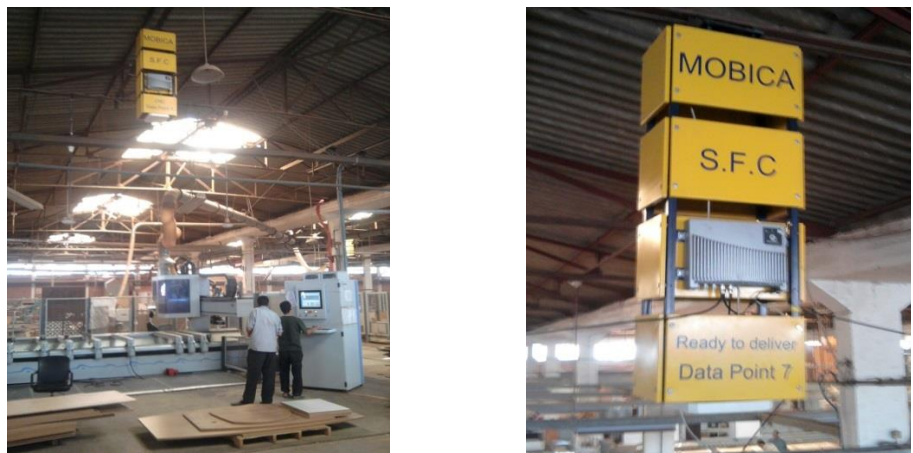

Fig.2- The data point Reader / antenna set

At the exit of the shop floor ,a product tag carrying the product ID ,is attached to the output product in preparation for the inventory of finished/semifinished product and customer delivery. The tag is read on the output gate of the shopfloor and information is transferred to the central ERP .

## Technical structure Decomposition

The structure of the SFC is illustrated in Fig.3. It is derived from the ISA 95 model (Enterprise-Control System Integration Standard, from ISA ,
https://www.isa.org/store/products/product-detail/?productId=116636 )

- The bottom layer is the physical RFID network which contains: RFID tags,antennas,readers,touch screens , and the SFC shop floor communication network

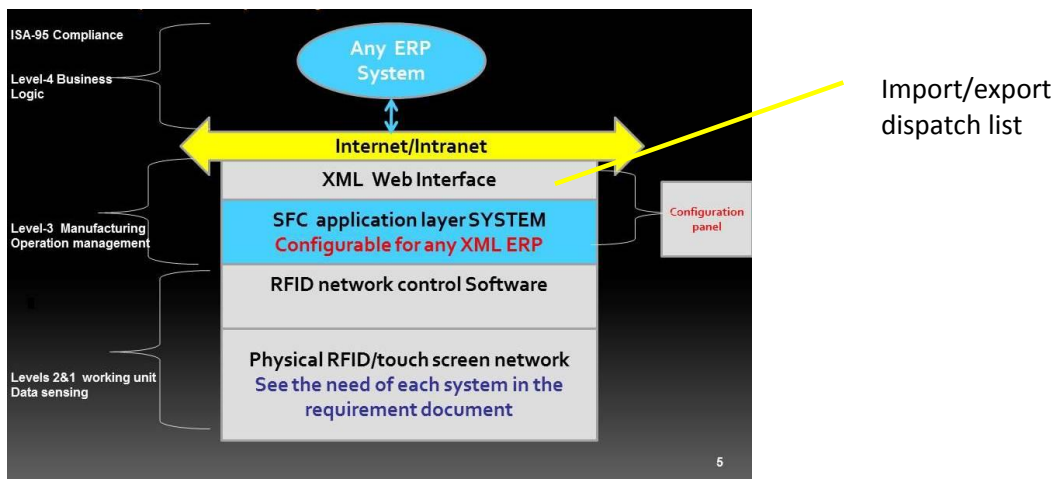

Fig.3- the structure of the SFC system

- The second layer upward is the network control middleware which is responsible for:
1. Identification of the tags IDs
2. Read/write the tags
3. Manage the data transfer between the data point and the reader
4. Manage the transfer between the readers network and the associated antennas ,to the SFC system
5. Manage the RF technology problems on the shop floor

- The SFC application layer is executing the SFC processes using Java-Eclipse environment

- The upper XML web interface, Configurable Integration with Production Planning System (ERP)-based on XML technology and the web services description language

Page **5** of **10**

## The economical and the standards constraints in the local market

These constraints impose:

1) The use of long range (to accommodate physical size of moving objects ) , passive RFID tags (up to 5 mts)  , these tags are low cost and allow a read/write scope of a circle with radius 3.6 mt.
2) The design uses  modern tag satisfying  the standard  EPC Class 1 Gen 2 which operates at UHF frequencies
3) For low cost tags the memory available is 512 bits . This restrict the amount of data stored in the tag . We decided to use Build Ticket tag that contains only the { Sales order(or internal make-to-stock order ) ,productID ,build ticket tag ID }. Any other information ,such as the routing , is obtained from the daily dispatch list stored in  the SFC data base.  .
4) The Frequency is UHF band, 865.7 – 867.5 MHz (Europe band) permitted in Egypt
5) The International low-level-reader-protocol LLRP is used to simplify developing standard APIs according to the requirements of this project.
6) The software is  based on the popular IDE Eclipse-Java. Eclipse is a Java-based, extensible open source development platform. By itself, it is simply a framework and a set of services for building applications from plug-in components
7) The SFC system does not  add any required user licenses to the back-end ERP. Most of the installed ERP base in Egypt are international packages with high cost/user license.

## Lessons learned

RFID is a tool to monitor pre-specified objects. Applying RFID requires first answering  the questions :

1. How can I improve the target business processes by having accurate and timely information . In our case , it is ERP interfacing to the shop floor using  RFID networks to provide feedback to the ERP and improve real-time shopfloor visibility
2. Given a factory/factories, define what the optimum distribution and types  of RFID tags, readers, and antennas on the shop floor, that allow real-time data reading and write ..In our case , we defined 7-data points for the shop floor shown in Fig.4. The communication  network of the  data points is shown in Fig.5.

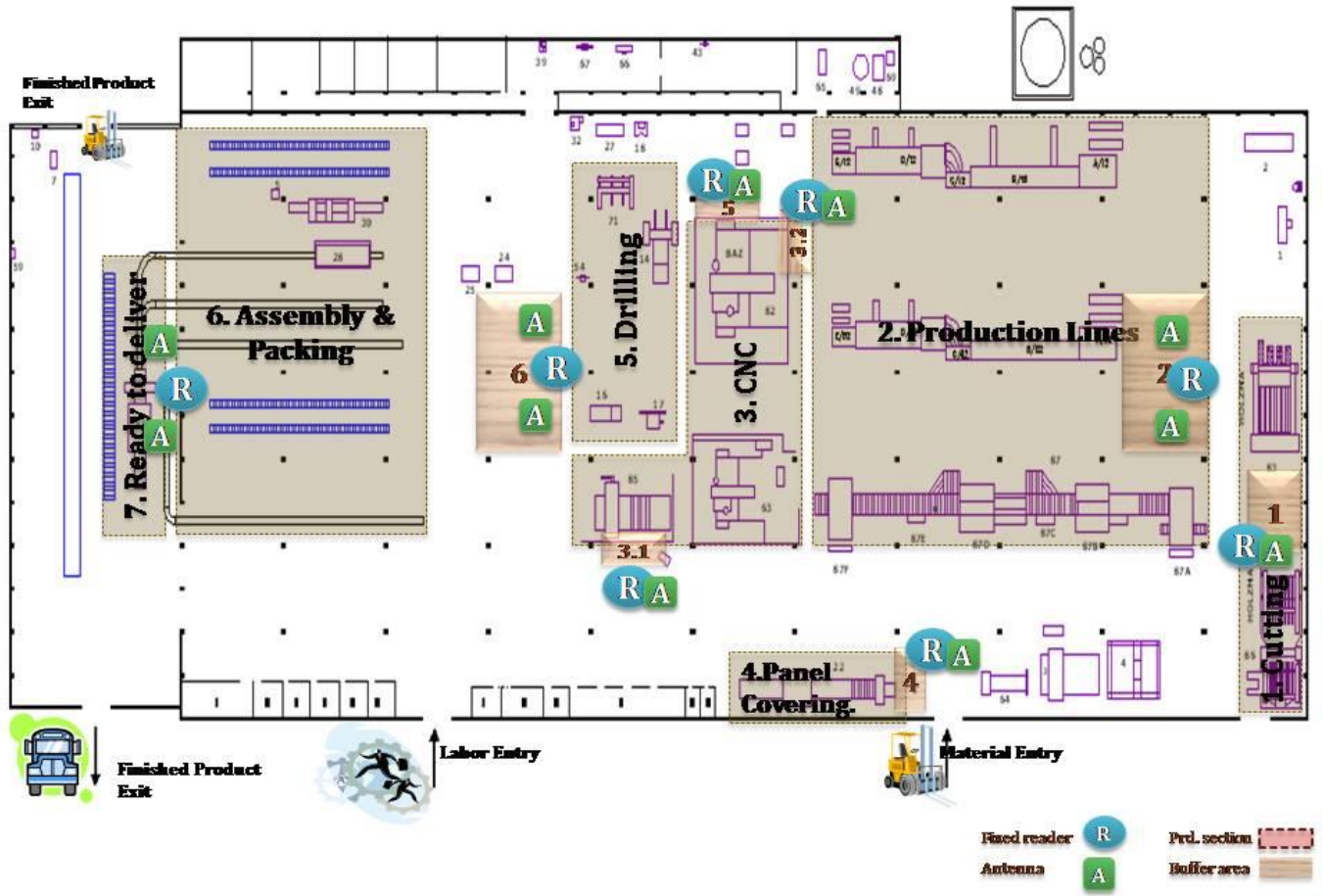

Fig.4 shop floor layout

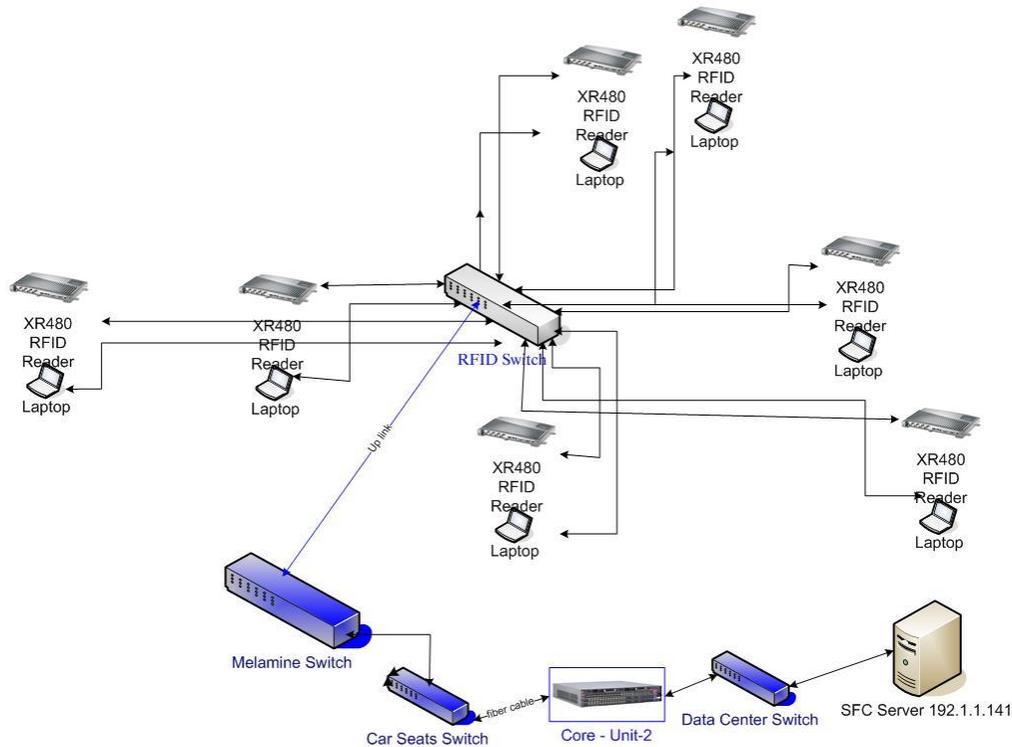

Fig.5- the shop floor network

3. What are the target objects to track
4. RFID eliminates or reduce the manual intervention in the business process cycle. A relevant question is therefore " Is there any serious risk by eliminating the manual intervention ?".
5. What kind of tags in terms of Read/write distance, frequency spectrum, interference sources , the conductivity of fixing surface
6. The level of monitored granularity . Is it the piece level ,which can be in discrete manufacturing, or at the material batch-level for flow – industry.
7. What is the location accuracy required for each item being tracked

The above problem handling questions are vital before starting any technical design. The next step is to start the design bottom-up according to the structure shown in Fig.3

Based on the answers , we can start moving upward to the network control of Fig.3.For the shopfloor network software , It is recommended using standard protocol " Low-Level-Reader-Protocol "(LLRP), this allows generating the required APIs according to the model of the SFC software. Most of the development iterations are between the SFC application and the RFID network control

# The project management stages

The project is implemented through the following work packages:

**WP1 Technology exploration and Requirements specifications**

**WP2 Product Design**

**WP3 Product Development and Implementation**

It includes

1. Developing the equipments specifications and ordering according to national rules
2. Installing the development environment using Eclipse/java environment
3. Development staff training
4. Product development and design tuning
5. Internal lab testing of the system

**WP4 Product Testing, Tuning and Validation**

It includes :

1. Equipments delivery
2. Installation at Mobica
3. Life pilot and further tuning
4. Launch the G0-life operation
5. Operation manual

**WP5 Dissemination and Product Commercialization Plan**

The marketing and commercialization plan is launched in March 2014 coinciding with The Ministry of Research conference held from 2/3--5/3 (national conference and exhibition on " connecting the Industry and the Research).

***Dr. Adel Ghannam*** , is teaching MIS at MSA university ,Cairo,Egypt and is a consultant in IT alignment with business strategy .He was the technical manager of the project published in this paper. You can contact him at adel.ghannam@isg-egypt.net

***Dr. Salwa Nassar*** , is professor of Computer systems at the Electronic Research Institute,Egypt, Cairo , you can contact her at   salwa.nassar79@gmail.com

***Dr. Hesham El Deeb*** , is professor of computer systems and president of the Electronic Research Institute , Egypt, Cairo, you can contact him at   heldeeb@mcit.gov.eg